\documentclass[twocolumn, superscriptaddress, amsmath,amssymb, prc aps]{revtex4-1}

\def\129Xe{$^{129}$Xe}
\def\Xe128{$^{128}$Xe}
\def\I127{$^{127}$I}

\usepackage{romanbar}
\usepackage{graphicx}
\usepackage{dcolumn}
\usepackage{bm}
\usepackage{array}
\usepackage{graphicx}  
\usepackage{siunitx}  
\usepackage{epsfig} 
\usepackage{subfigure}
\usepackage{lineno}  
\usepackage{amsmath,amssymb}
\usepackage{comment}
\usepackage{multirow}
\usepackage{dcolumn}
\usepackage{bm}
\usepackage{setspace}
\usepackage{multibib}
\includecomment{printrobustnotes}
\includecomment{printallnotes}
\usepackage{xspace}
\usepackage{color}
\usepackage{hyperref} 
\begin{document}

\title{Search for inelastic WIMP-iodine scattering with COSINE-100}
\author{G.~Adhikari}
\affiliation{Department of Physics and Wright Laboratory, Yale University, New Haven, CT 06520, USA}
\author{N.~Carlin}
\affiliation{Physics Institute, University of S\~{a}o Paulo, 05508-090, S\~{a}o Paulo, Brazil}
\author{J.~Y.~Cho}
\affiliation{Center for Underground Physics, Institute for Basic Science (IBS), Daejeon 34126, Republic of Korea}
\author{J.~J.~Choi}
\affiliation{Department of Physics and Astronomy, Seoul National University, Seoul 08826, Republic of Korea} 
\affiliation{Center for Underground Physics, Institute for Basic Science (IBS), Daejeon 34126, Republic of Korea}
\author{S.~Choi}
\affiliation{Department of Physics and Astronomy, Seoul National University, Seoul 08826, Republic of Korea} 
\author{A.~C.~Ezeribe}
\affiliation{Department of Physics and Astronomy, University of Sheffield, Sheffield S3 7RH, United Kingdom}
\author{L.~E.~Fran{\c c}a}
\affiliation{Physics Institute, University of S\~{a}o Paulo, 05508-090, S\~{a}o Paulo, Brazil}
\author{C.~Ha}
\affiliation{Department of Physics, Chung-Ang University, Seoul 06973, Republic of Korea}
\author{I.~S.~Hahn}
\affiliation{Center for Exotic Nuclear Studies, Institute for Basic Science (IBS), Daejeon 34126, Republic of Korea}
\affiliation{Department of Science Education, Ewha Womans University, Seoul 03760, Republic of Korea} 
\affiliation{IBS School, University of Science and Technology (UST), Daejeon 34113, Republic of Korea}
\author{S.~J.~Hollick}
\affiliation{Department of Physics and Wright Laboratory, Yale University, New Haven, CT 06520, USA}
\author{E.~J.~Jeon}
\affiliation{Center for Underground Physics, Institute for Basic Science (IBS), Daejeon 34126, Republic of Korea}
\affiliation{IBS School, University of Science and Technology (UST), Daejeon 34113, Republic of Korea}
\author{H.~W.~Joo}
\affiliation{Department of Physics and Astronomy, Seoul National University, Seoul 08826, Republic of Korea} 
\author{W.~G.~Kang}
\affiliation{Center for Underground Physics, Institute for Basic Science (IBS), Daejeon 34126, Republic of Korea}
\author{M.~Kauer}
\affiliation{Department of Physics and Wisconsin IceCube Particle Astrophysics Center, University of Wisconsin-Madison, Madison, WI 53706, USA}
\author{B.~H.~Kim}
\affiliation{Center for Underground Physics, Institute for Basic Science (IBS), Daejeon 34126, Republic of Korea}
\author{H.~J.~Kim}
\affiliation{Department of Physics, Kyungpook National University, Daegu 41566, Republic of Korea}
\author{J.~Kim}
\affiliation{Department of Physics, Chung-Ang University, Seoul 06973, Republic of Korea}
\author{K.~W.~Kim}
\affiliation{Center for Underground Physics, Institute for Basic Science (IBS), Daejeon 34126, Republic of Korea}
\author{S.~H.~Kim}
\affiliation{Center for Underground Physics, Institute for Basic Science (IBS), Daejeon 34126, Republic of Korea}
\author{S.~K.~Kim}
\affiliation{Department of Physics and Astronomy, Seoul National University, Seoul 08826, Republic of Korea}
\author{S.~W.~Kim}
\affiliation{Center for Underground Physics, Institute for Basic Science (IBS), Daejeon 34126, Republic of Korea}
\author{W.~K.~Kim}
\affiliation{IBS School, University of Science and Technology (UST), Daejeon 34113, Republic of Korea}
\affiliation{Center for Underground Physics, Institute for Basic Science (IBS), Daejeon 34126, Republic of Korea}
\author{Y.~D.~Kim}
\affiliation{Center for Underground Physics, Institute for Basic Science (IBS), Daejeon 34126, Republic of Korea}
\affiliation{Department of Physics, Sejong University, Seoul 05006, Republic of Korea}
\affiliation{IBS School, University of Science and Technology (UST), Daejeon 34113, Republic of Korea}
\author{Y.~H.~Kim}
\affiliation{Center for Underground Physics, Institute for Basic Science (IBS), Daejeon 34126, Republic of Korea}
\affiliation{Korea Research Institute of Standards and Science, Daejeon 34113, Republic of Korea}
\affiliation{IBS School, University of Science and Technology (UST), Daejeon 34113, Republic of Korea}
\author{Y.~J.~Ko}
\affiliation{Center for Underground Physics, Institute for Basic Science (IBS), Daejeon 34126, Republic of Korea}
\author{D.~H.~Lee}
\affiliation{Department of Physics, Kyungpook National University, Daegu 41566, Republic of Korea}
\author{E.~K.~Lee}
\affiliation{Center for Underground Physics, Institute for Basic Science (IBS), Daejeon 34126, Republic of Korea}
\author{H.~Lee}
\affiliation{IBS School, University of Science and Technology (UST), Daejeon 34113, Republic of Korea}
\affiliation{Center for Underground Physics, Institute for Basic Science (IBS), Daejeon 34126, Republic of Korea}
\author{H.~S.~Lee}
\email{hyunsulee@ibs.re.kr}
\affiliation{Center for Underground Physics, Institute for Basic Science (IBS), Daejeon 34126, Republic of Korea}
\affiliation{IBS School, University of Science and Technology (UST), Daejeon 34113, Republic of Korea}
\author{H.~Y.~Lee}
\affiliation{Center for Exotic Nuclear Studies, Institute for Basic Science (IBS), Daejeon 34126, Republic of Korea}
\author{I.~S.~Lee}
\email{islee@ibs.re.kr}
\affiliation{Center for Underground Physics, Institute for Basic Science (IBS), Daejeon 34126, Republic of Korea}
\author{J.~Lee}
\affiliation{Center for Underground Physics, Institute for Basic Science (IBS), Daejeon 34126, Republic of Korea}
\author{J.~Y.~Lee}
\affiliation{Department of Physics, Kyungpook National University, Daegu 41566, Republic of Korea}
\author{M.~H.~Lee}
\affiliation{Center for Underground Physics, Institute for Basic Science (IBS), Daejeon 34126, Republic of Korea}
\affiliation{IBS School, University of Science and Technology (UST), Daejeon 34113, Republic of Korea}
\author{S.~H.~Lee}
\affiliation{IBS School, University of Science and Technology (UST), Daejeon 34113, Republic of Korea}
\affiliation{Center for Underground Physics, Institute for Basic Science (IBS), Daejeon 34126, Republic of Korea}
\author{S.~M.~Lee}
\affiliation{Department of Physics and Astronomy, Seoul National University, Seoul 08826, Republic of Korea} 
\author{Y.~J.~Lee}
\affiliation{Department of Physics, Chung-Ang University, Seoul 06973, Republic of Korea}
\author{D.~S.~Leonard}
\affiliation{Center for Underground Physics, Institute for Basic Science (IBS), Daejeon 34126, Republic of Korea}
\author{N.~T.~Luan}
\affiliation{Department of Physics, Kyungpook National University, Daegu 41566, Republic of Korea}
\author{B.~B.~Manzato}
\affiliation{Physics Institute, University of S\~{a}o Paulo, 05508-090, S\~{a}o Paulo, Brazil}
\author{R.~H.~Maruyama}
\affiliation{Department of Physics and Wright Laboratory, Yale University, New Haven, CT 06520, USA}
\author{R.~J.~Neal}
\affiliation{Department of Physics and Astronomy, University of Sheffield, Sheffield S3 7RH, United Kingdom}
\author{J.~A.~Nikkel}
\affiliation{Department of Physics and Wright Laboratory, Yale University, New Haven, CT 06520, USA}
\author{S.~L.~Olsen}
\affiliation{Center for Underground Physics, Institute for Basic Science (IBS), Daejeon 34126, Republic of Korea}
\author{B.~J.~Park}
\affiliation{IBS School, University of Science and Technology (UST), Daejeon 34113, Republic of Korea}
\affiliation{Center for Underground Physics, Institute for Basic Science (IBS), Daejeon 34126, Republic of Korea}
\author{H.~K.~Park}
\affiliation{Department of Accelerator Science, Korea University, Sejong 30019, Republic of Korea}
\author{H.~S.~Park}
\affiliation{Korea Research Institute of Standards and Science, Daejeon 34113, Republic of Korea}
\author{J.~C.~Park}
\affiliation{Department of Physics and IQS, Chungnam National University, Daejeon 34134, Republic of Korea}
\author{K.~S.~Park}
\affiliation{Center for Underground Physics, Institute for Basic Science (IBS), Daejeon 34126, Republic of Korea}
\author{S.~D.~Park}
\affiliation{Department of Physics, Kyungpook National University, Daegu 41566, Republic of Korea}
\author{R.~L.~C.~Pitta}
\affiliation{Physics Institute, University of S\~{a}o Paulo, 05508-090, S\~{a}o Paulo, Brazil}
\author{H.~Prihtiadi}
\affiliation{Department of Physics, Universitas Negeri Malang, Malang 65145, Indonesia}
\author{S.~J.~Ra}
\affiliation{Center for Underground Physics, Institute for Basic Science (IBS), Daejeon 34126, Republic of Korea}
\author{C.~Rott}
\affiliation{Department of Physics, Sungkyunkwan University, Suwon 16419, Republic of Korea}
\affiliation{Department of Physics and Astronomy, University of Utah, Salt Lake City, UT 84112, USA}
\author{K.~A.~Shin}
\affiliation{Center for Underground Physics, Institute for Basic Science (IBS), Daejeon 34126, Republic of Korea}
\author{D.~F.~F.~S. Cavalcante}
\affiliation{Physics Institute, University of S\~{a}o Paulo, 05508-090, S\~{a}o Paulo, Brazil}
\author{A.~Scarff}
\affiliation{Department of Physics and Astronomy, University of Sheffield, Sheffield S3 7RH, United Kingdom}
\author{M.~K.~Son}
\affiliation{Department of Physics and IQS, Chungnam National University, Daejeon 34134, Republic of Korea}
\author{N.~J.~C.~Spooner}
\affiliation{Department of Physics and Astronomy, University of Sheffield, Sheffield S3 7RH, United Kingdom}
\author{L.~T.~Truc}
\affiliation{Department of Physics, Kyungpook National University, Daegu 41566, Republic of Korea}
\author{L.~Yang}
\affiliation{Department of Physics, University of California San Diego, La Jolla, CA 92093, USA}
\author{G.~H.~Yu}
\affiliation{Department of Physics, Sungkyunkwan University, Suwon 16419, Republic of Korea}
\affiliation{Center for Underground Physics, Institute for Basic Science (IBS), Daejeon 34126, Republic of Korea}
\collaboration{COSINE-100 Collaboration}
\date{\today}

\begin{abstract}
		We report the results of a search for inelastic scattering of weakly interacting massive particles (WIMPs) off $^{127}$I nuclei using NaI(Tl) crystals with a data exposure of 97.7\,kg$\cdot$years from the COSINE-100 experiment. 
		The signature of inelastic WIMP-\I127 scattering is a nuclear recoil accompanied by a 57.6\,keV $\gamma$-ray from the prompt deexcitation, producing a more energetic signal compared to the typical WIMP nuclear recoil signal. 
		We found no evidence for this inelastic scattering signature and set a 90 $\%$ confidence level upper limit on the WIMP-proton spin-dependent, inelastic scattering cross section of $1.2 \times 10^{-37} {\rm cm^{2}}$ at the WIMP mass 500\,${\rm GeV/c^{2}}$.
\end{abstract}

\maketitle

\section{Introduction} 

An abundance of astronomical observations indicates the existence of invisible dark matter in the Universe~\cite{Clowe:2006eq,Aghanim:2018eyx}.
Among the various candidates for the particle dark matter, the weakly interacting massive particle~(WIMP)~\cite{PhysRevLett.39.165,Goodman:1984dc} is a prominent candidate that is well motivated by theoretical models beyond the standard model of particle physics~\cite{Jungman:1995df,Peskin:2014wta,Throm:2019vhd,Hooper:2007qk,Chen:2014wua}.
Several direct~\cite{Billard:2021uyg} and indirect~\cite{Conrad:2017pms} detection experiments, as well as collider production experiments~\cite{dmcol}, have tested this hypothesis, but no clear evidence has been yet observed~\cite{ParticleDataGroup:2022pth}. 

Most direct detection experiments have focused on WIMP-nucleus elastic scatterings reactions, which result in nuclear recoils with energy spectra that fall exponentially~\cite{Schumann:2019eaa}. These spectra typically have a maximum energy of less than 100\,keVnr (kilo-electron-volt nuclear recoil energy). In case of scintillation or ionization measurement detectors, nuclear recoil energies are quenched to less than 10\,keVee (kilo-electron-volt electron-equivalent energy) due to the detector responses~\cite{Joo:2018hom,Collar:2019ihs,Goetzke:2016lfg,DarkSide:2021bnz,Ramanathan:2017dfn,Collar:2021fcl,Bonhomme:2022lcz}.
Considering the exponentially falling recoil spectra, understanding low-energy events close to the threshold is beneficial for WIMP elastic scattering~\cite{DarkSide-50:2022qzh,LZ:2022lsv,XENON:2023cxc,ParticleDataGroup:2022pth}

On the other hand, when WIMP scatters from a target nucleus in a detector, it may excite the nucleus to a higher energy state. In this so called WIMP-nucleus inelastic scattering scenario~\cite{ELLIS1988375,Vergados:2013raa,PhysRevC.102.035501}, the subsequent deexcitation of the nucleus emits a characteristic $\gamma$-ray that is detected in addition to the nuclear recoil energy transferred from the WIMP-nucleus interaction. In this scenario, the inelastic interaction produces a relatively high-energy signature that is well above the detector's threshold. 
Since the nuclear excitation in inelastic scattering is a spin-dependent interaction, the observation of WIMP inelastic scattering would provide insight into spin properties of WIMPs. 
In the model of Ref.~\cite{Arcadi:2019hrw}, inelastic scattering channels can have better sensitivity for some ranges of WIMP model parameters.

Previous searches for  WIMP inelastic scattering have focused on excited states of \I127~\cite{FUSHIMI1994400} or \129Xe ~\cite{xmass:20191,PhysRevD.96.022008,PhysRevD.103.063028}.
In the case of \129Xe, the inelastic nuclear excitation of the 39.6\,keV state relies on the WIMP-neutron spin-dependent interaction.  
With a 26\,\% natural abundance of \129Xe, the exclusion limits on the WIMP-neutron inelastic scattering channels have been obtained~\cite{xmass:20191,PhysRevD.103.063028}. 

The \I127 nucleus has complementary advantages over the \129Xe nucleus; it has almost 100\,\% natural abundance and odd proton numbers sensitive to WIMP-proton spin-dependent interaction. 
The first excited  level of \I127 is a 7/2$^+$ state of 57.6\,keV above the 5/2$^+$ ground state, with a half-life of 1.9\,ns~\cite{Vergados:2013raa}. 
Therefore, the WIMP-\I127 inelastic interaction signal would be a combination of the nuclear recoil and the 57.6\,keV deexcitation energies. The ELEGANTS-V NaI(Tl) experiment~\cite{FUSHIMI1994400} reported limits on the rate of events rate with energies in this region. 
In this paper, we report a search for the WIMP-\I127 inelastic scattering events from the COSINE-100 NaI(Tl) crystal detectors in the signal region of 35--85\,keVee. 

\section{Experiment} 

\begin{figure}[!htb]
\begin{center}
\includegraphics[width=0.49\textwidth]{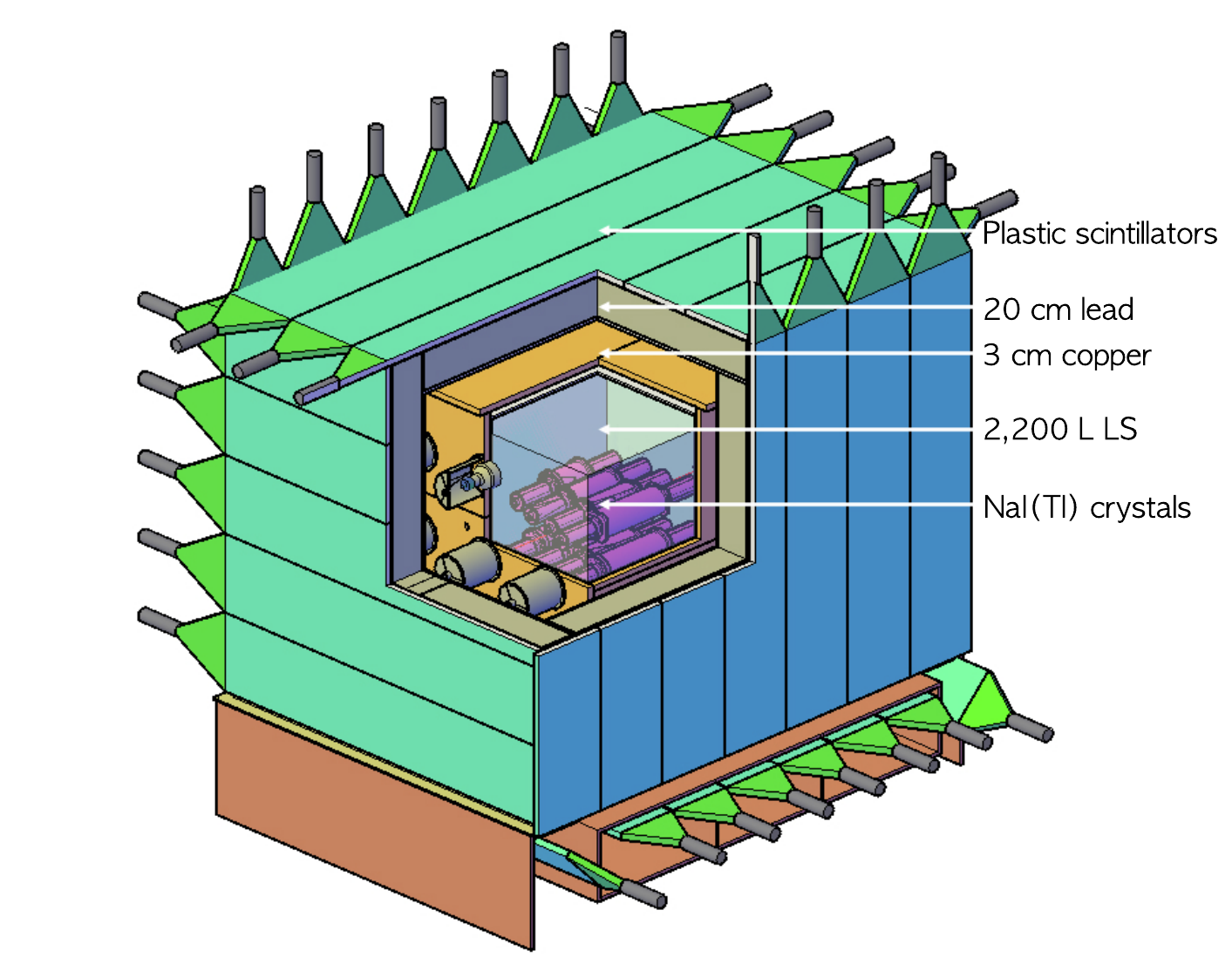} 
\caption{Schematic of the COSINE-100 detector. The NaI(Tl) detectors are immersed in the 2,200\,l LAB-LS that is surrounded by layers of copper and lead shielding.}
\label{fig_det}
\end{center}
\end{figure}

The COSINE-100 experiment~\cite{Adhikari:2017esn} is located at the Yangyang underground laboratory in South Korea, with an overburden of approximately 700\,m~\cite{Prihtiadi:2017inr}.
The COSINE-100 detector, as shown in Fig.~\ref{fig_det}, consists of an array of eight ultra-pure NaI(Tl) crystals~\cite{cosinebg,cosinebg2}, total weight of 106\,kg.
The NaI(Tl) crystal assemblies are immersed in an active veto detector comprised of 2,200\,l of linear alkylbenzene~(LAB)-based liquid scintillator~(LS) that attenuates or tags the influence of radioactive backgrounds observed by crystals~\cite{Adhikari:2020asl}.
The LAB-LS is surrounded by a 3\,cm-thick layer of oxygen-free copper, a 20\,cm thick lead shield, and plastic scintillator pannels that tag and veto cosmic ray muons~\cite{Prihtiadi:2017inr,Prihtiadi:2020yhz}.
Each crystal is optically coupled to two photomultiplier tubes (PMTs).

An event is triggered when a signal corresponding to one or more photoelectrons occurs in both PMTs of a crystal that are within a 200\,ns time window.
If at least one crystal satisfies the trigger condition, signals from all crystals and the LAB-LS are recorded.
The signals from the crystal PMTs are 8\,$\mu$s long waveforms that start 2.4\,$\mu$s earlier than the trigger position, and are processed by 500\,MHz flash analog-to-digital converters (FADCs). 
In addition to the 5$^{\textrm{th}}$ stage dynode readouts for high energies (50--3000\,keVee), the low-energy (0--100\,keVee) anode readouts are stored by the FADCs.  
The LAB-LS and plastic scintillator signals are processed by charge-sensitive FADCs.
Muon events are triggered by coincident signals from at least two plastic scintillator panels.
A trigger and clock board reads the trigger information from individual boards and generates a global trigger and time synchronizations for all of the modules.
Details of the COSINE-100 data acquisition system are described in Ref.~\cite{Adhikari:2018fpo}. 

The analysis presented here utilizes data from October 2016 to July 2018, corresponding to 1.7 years of exposure, which were used for our first annual modulation search~\cite{Adhikari:2019off} and the model-dependent WIMP dark matter search that was based on the shape of the energy spectra~\cite{COSINE-100:2021xqn}.
During the 1.7-year data-taking period, no significant environmental anomalies or unstable detector performance were observed.
Three of the eight crystals are excluded from this analysis due to their high background, high noise, and low light yield, resulting in a total effective mass of 61.3\,kg~\cite{Adhikari:2019off,COSINE-100:2021xqn}.

\section{WIMP-\I127 inelastic scattering signals}

An inelastic scattering event that occurs in \I127 results in the nuclear recoil together with the  emission of the 57.6\,keV $\gamma$-ray from the deexcitation. 
The nuclear recoil event rate depends on the velocity distribution of the WIMPs in the galatic dark matter halo and the nuclear form factor for the spin dependent interaction. 
The differential nuclear recoil rate per unit energy of the nuclear components is described as follows~\cite{Zyla:2020zbs},
\begin{eqnarray}
  \frac{d{R}}{dE_{\rm nr}} = \frac{\rho_{\chi}}{2m_{\chi}\mu^{2}}\sigma\int^{v_{\rm max}}_{v_{\rm min}}d^{3}f({\bf v},t) ,
 \label{eq:NuclearRecoil}
\end{eqnarray}
where ${R}$ is the event rate per unit target mass and unit time, $E_{\rm nr}$ is the nuclear recoil energy, $\rho_{\chi}$ is the local dark matter density, $m_{\chi}$ and $\mu$ are the WIMP mass and the reduced mass of WIMP and \I127 nucleus, respectively, and $\sigma$ is the WIMP-nucleus scattering cross section for the inelastic interaction. 
The integral is performed from the minimum velocity $v_{\rm min}$ required to excite a \I127 nucleus to the maximum velocity $v_{\rm max}$, which is the same as the galactic escape velocity $v_{\rm esc}$. 
Because of the excitation energy $E_{\rm ex}$, $v_{\rm min}$ is increased from the minimum velocity of the typical nuclear recoil ($v_0$) as follows, 
\begin{eqnarray}
		v_{\rm min} = v_0 + \frac{v^{2}_{\rm thr}}{4v_0}, 
 \label{eq:vmin}
\end{eqnarray}
where $v_0 = \sqrt{\frac{m_{\rm target}E_{\rm nr}}{2\mu^{2}}}$, $m_{\rm target}$ $=$ the mass of \I127, and $v^{2}_{\rm thr} = \frac{2\ E_{\rm ex}}{\mu}$.

The cross section can be expressed in terms of the cross section of the WIMP-proton spin-dependent interaction $\sigma_{\rm p}$ by 
\begin{eqnarray}
  \sigma = \frac{4}{3}\frac{\pi}{2J+1}\left(\frac{\mu}{\mu_{\rm p}}\right)^{2}S(E_{\rm nr})\sigma_{\rm p},
 \label{eq:sigma}
\end{eqnarray}
where $J = 5/2$ is the spin of the \I127 ground-state, $\mu_{\rm p}$ is the reduced mass of the WIMP and the proton, and $S(E_{\rm nr})$ is the nuclear form factor of inelastic WIMP-\I127 scattering. We use a recent calculation of $S(E_{\rm nr})$ for the \I127 inelastic interaction in Ref.~\cite{PhysRevC.102.035501}. 
We assume the standard halo model of the WIMP velocity distribution, $f({\bf v}, t)$~\cite{Lewin:1995rx,Freese:2012xd},
\begin{eqnarray}
		f({\bf v},t) = \begin{cases}
  	\frac{1}{N_\mathrm{esc}}&\left(\frac{3}{2\pi\sigma_v^2}\right)^{3/2}e^{-3[{\bf v}+{\bf v}_\mathrm{E}(t)]^2/2\sigma_v^2},\\
		&\mbox{for }\left|{\bf v}+{\bf v}_\mathrm{E}(t)\right|<v_{\rm esc}\\
	0,&\mbox{otherwise,}
  \end{cases}
\label{eq:vdist}
\end{eqnarray}
where $N_\mathrm{esc}$ is a normalization constant, ${\bf v}_\mathrm{E}$ is the Earth velocity relative to the WIMP dark matter and $\sigma_v$ is the velocity dispersion. The standard halo model parameterization is used with the local dark matter density $\rho_{\chi} = 0.3$\,GeV/$c^{2}$/cm$^3$, $v_\mathrm{E}$ = 232\,km/s, $\sqrt{2/3}\sigma_v$ = 220\,km/s, and $v_{\rm esc}$ = 544\,km/s~\cite{Smith:2006ym}. 

\begin{figure}[!htb]
\begin{center}
\includegraphics[width=0.49\textwidth]{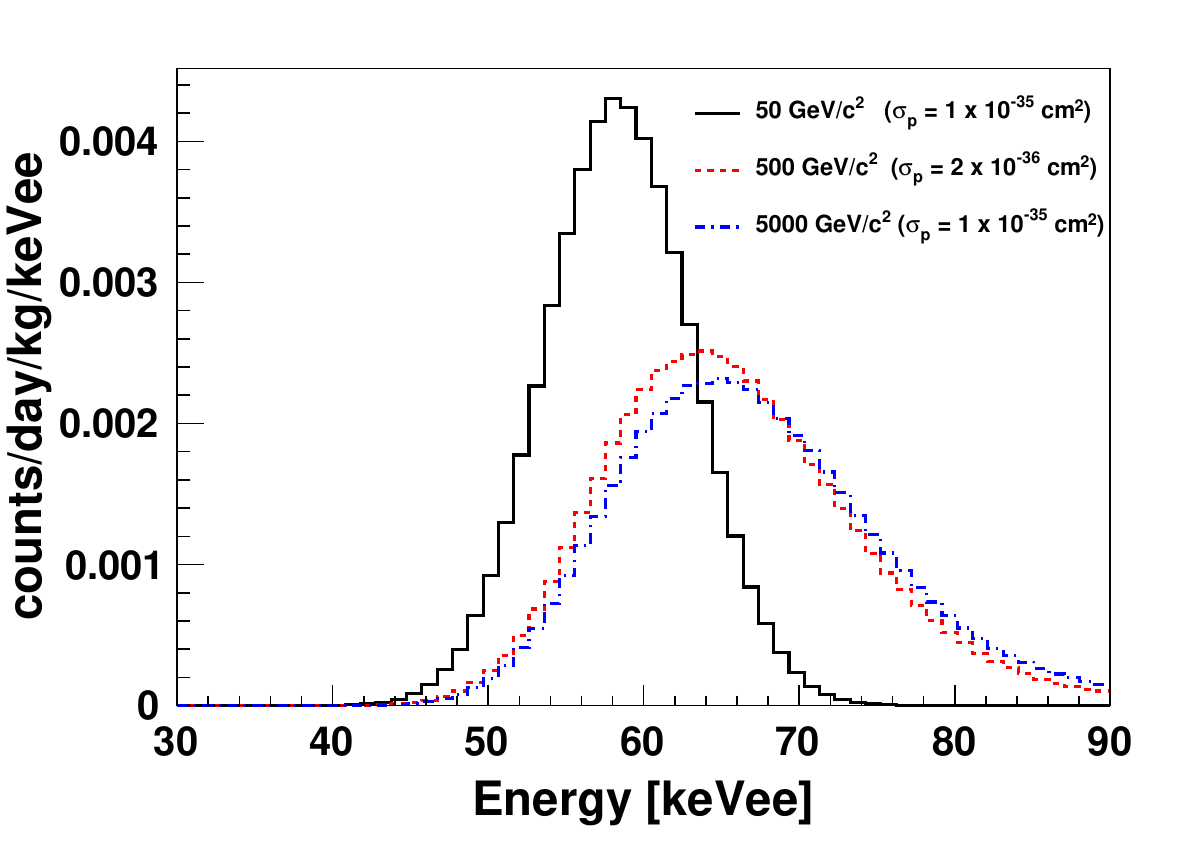} 
\caption{The expected energy spectra for  WIMP-\I127 inelastic scattering in the COSINE-100 detector are shown for WIMP masses of 50~${\rm GeV/c^{2}}$, 500~${\rm GeV/c^{2}}$, and 5,000~${\rm GeV/c^{2}}$.  The spectra include the nuclear recoil energy and the deexcitation energy.} 
\label{fig:sigMC}
\end{center}
\end{figure}

Because of the short half-life of 1.9\,ns of the  57.6\,keV excited \I127 state, the energy deposited in the detector will be the sum of the nuclear recoil energy and the deexcitation energy of 57.6\,keV. 
Since the nuclear recoil energy is quenched to the visible electron-equivalent energy with approximately 5--10\,\% levels~\cite{Joo:2018hom}, the visible energy is expressed as follows,
\begin{eqnarray}
E_{\rm vis} = f(E_{\rm nr}) \times E_{\rm nr} + E_{\rm ex}.
 \label{eq:visibleenergy}
\end{eqnarray}
Here, $f(E_{\rm nr})$ is the energy-dependent quenching factor for nuclear recoils.
In this search, we use the measured quenching factor of iodine from Ref.~\cite{Joo:2018hom} with an empirical model described in Ref.~\cite{Ko:2019enb}.
Figure~\ref{fig:sigMC} shows the simulated inelastic scattering energy spectra for the WIMP masses of 50\,GeV/c$^2$, 500\,GeV/c$^{2}$, and 5,000 GeV/c$^2$ based on Eq.~\ref{eq:NuclearRecoil}. In these energy spectra, energy resolutions of individual crystal detectors~\cite{cosinebg2} are taken into account. 
The nuclear recoil energy is more relevant for high-mass WIMPs, and long tails of the energy spectra at high energy are evident.

\section{Data analysis}

\subsection{Event selection}
In order to suppress cosmic-ray  muon-induced events, the crystal hit events that are coincident with muon candidate events in the muon detector~\cite{Prihtiadi:2017inr,Prihtiadi:2020yhz} within 30\,ms are rejected.
Additionally, we require that the leading edges of the trigger pulses start later than 2.0\,$\mu$s after the start of the recording, that the waveforms from the hit crystal contain more than two single photoelectrons, and that the integral waveform area below the baseline does not exceed a certain limit.
These criteria reject muon-induced phosphor events and electronic interference noise events.

A multiple-hit event is one in which more than one crystal has a signal containing more than four photoelectrons in an 8\,$\mu$s time window or has an LS signal above an 80\,keVee threshold within 4\,$\mu$s of the hit crystal~\cite{Adhikari:2020asl}. A single-hit event is classified as one where only one crystal has a hit, and none of the other detectors meets the above criteria. In this analysis, only single-hit events are used. 

In the signal region around 57.6\,keVee energy, there is a few-percent-level contamination from the high energy tail of PMT-induced noise events. However, these events are efficiently rejected with the boosted decision tree-based discriminant, as described in Refs.~\cite{Adhikari:2020xxj,COSINE-100:2021poy}. The selection efficiencies for the scintillation events are estimated from calibration data using a $^{60}$Co source, which are Compton scattered events of 1.17\,MeV and 1.33\,MeV $\gamma$s, to be more than 99\,\% in the 35--85\,keVee signal region~\cite{Adhikari:2020xxj}

\subsection{Backgrounds}
\begin{figure}[!htb]
\begin{center}
\includegraphics[width=0.49\textwidth]{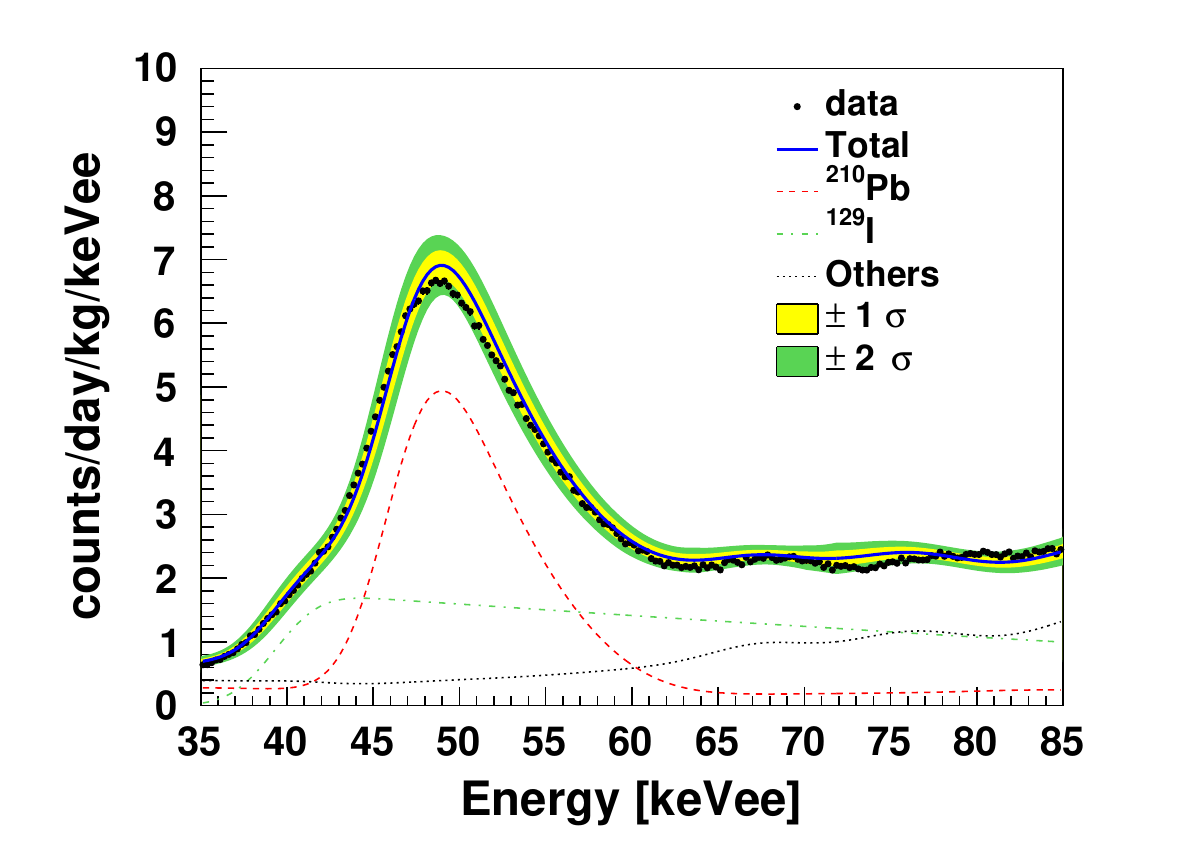} 
\caption{Measured energy spectrum in the ROI of crystal 4 (black points) and its background model (blue solid line) with the 68\% (yellow band) and 95\% (green band) confidence intervals are presented. The expected contributions to the background from  $^{210}$Pb (red dashed line), $^{129}$I (green dotted-dashed line), and other components (black dotted line) are indicated. In the 50--70\,keVee energy region, $^{210}$Pb and $^{129}$I are the dominant background components.}
\label{fig:background}
\end{center}
\end{figure}

Geant4~\cite{Agostinelli:2002hh}-based simulations are used to understand the contribution of each background component~\cite{cosinebg2}. 
The fraction of each of these  is determined from a simultaneous fit to the single-hit and multiple-hit events. 
For the single-hit data, we consider the fit energy range between 35 and 85\,keVee that covers the dominant inelastic signal range of 50--70\,keVee. 
Figure~\ref{fig:background} presents the crystal 4 energy spectrum in the results from the Ref.~\cite{cosinebg2} background model superimposed.  
In this ROI, the dominant backgrounds are the internal $^{210}$Pb, which  produces a 46.5\,keV $\gamma$-ray  with a few keV Auger electrons, and $^{129}$I, which emits a 39.6\,keV $\gamma$-ray and a $\beta$ particle.  
Table~\ref{table:Background} presents the expected background composition and the observed data in the 50--70\,keVee energy region of the single-hit events for the five crystals. As shown in this table, the observed data agree well with the sum of the expected backgrounds.

\begin{table*}[t!]
  \caption{The background expectations and observed data in the 50--70\,keVee energy range for single hits of the 1.7\,years COSINE-100 exposure are summarized, with only statistical uncertainties being considered. }
  \small
  \begin{center}
    \renewcommand{\arraystretch}{1.5}
    \begin{tabular}{ c c c c c c c }
      \hline \hline
      \multicolumn{2}{c}{}      & crystal 2 & crystal 3 & crystal 4 & crystal 6 & crystal 7  \\ \hline
      \multirow{3}{*}{Internal}    & $^{210}$Pb &  424500 $\pm$ 2400 & 155800 $\pm$ 4200 & 290000 $\pm$ 9100 & 486300 $\pm$ 8200 & 420900 $\pm$ 5100    \\ 
                                   & $^{40}$K &  8400 $\pm$ 200    & 3700 $\pm$ 100    & 8800  $\pm$ 100   & 2100 $\pm$ 200    & 2300 $\pm$ 200   \\
                                   &  Others & 100 $\pm$ 10      & 100 $\pm$ 10      & 100    $\pm$ 10      & 100 $\pm$ 11      & 100 $\pm$ 10      \\ \hline
      \multirow{3}{*}{External}    & $^{238}$U &  87200 $\pm$ 1900  & 68600 $\pm$ 1600  & 24900 $\pm$ 2100  & 71100 $\pm$ 1600  & 67300 $\pm$ 1700 \\ 
                                   & $^{228}$Th&  11600 $\pm$ 2500  & 17500 $\pm$ 1900  & 38700 $\pm$ 2800  & 34300 $\pm$ 2400  & 29300 $\pm$ 2100    \\
                                   & Others &  6400 $\pm$ 600    & 3400 $\pm$ 600    & 13400 $\pm$ 1000  & 12200 $\pm$ 900   & 8100 $\pm$ 1200 \\\hline
      \multirow{3}{*}{Cosmogenic}  & $^{129}$I &   157700 $\pm$ 3200 & 148200 $\pm$ 3300 & 331000 $\pm$ 5300 & 230300 $\pm$ 4200 & 313100 $\pm$ 5100  \\ 
                                   & $^{127m}$Te &  1900 $\pm$ 300    & 2900 $\pm$ 100    & 2100 $\pm$ 100    & 800 $\pm$ 100     & 800 $\pm$ 100 \\
                                   & Others  &  100 $\pm$ 8        & 2500 $\pm$ 280    & 32500 $\pm$ 900   & 18100 $\pm$ 1600  & 13600 $\pm$ 1600   \\ 
																	 \hline\hline
      \multicolumn{2}{c}{Total (expected)}    &   707900 $\pm$ 5100 & 402700 $\pm$ 5900 & 742800 $\pm$ 9800& 855300 $\pm$ 9800 & 855500 $\pm$ 8000    \\ \hline
      \multicolumn{2}{c}{Data}                 & 716352        & 410655     & 746285     & 856789  &    864034       \\ \hline\hline
    \end{tabular}
  \end{center}
  \label{table:Background}
\end{table*}

We consider various sources of systematic uncertainties in the background and signal models. Errors associated with the energy resolution, the energy scale, and the background modeling technique are accounted for in the shapes of the signal and background probability density functions, as well as in rate changes as described in Ref.~\cite{COSINE-100:2021xqn}. These quantities are allowed to vary within their uncertainties as nuisance parameters in the data fit used to extract the signal. 
The largest systematic uncertainty comes from the error associated with the $^{210}$Pb background modeling, which is due to its dominant contribution and its substantial shape change in the ROI as shown in Fig.~\ref{fig:background}. 
This error includes uncertainties in the energy resolution, energy scale, and depth profiles of $^{210}$Pb on the surface of the NaI(Tl) crystals, which were studied with a $^{222}$Rn contaminated crystal~\cite{Yu:2020ntl} and were varied within their uncertainties.

\subsection{Signal fit}
To search for evidence of the WIMP-\I127 inelastic scattering signals, a Bayesian approach with a likelihood function based on Poisson probability, described in Ref.~\cite{COSINE-100:2021xqn}, is used.
The likelihood fit is applied to the measured single-hit energy spectra between 35 and 85\,keVee for several WIMP masses. Each crystal is fitted with a crystal-specific background model and a crystal-correlated WIMP signal for the combined fit by multiplying the five crystals’ likelihoods. The means and uncertainties for background components, which are determined from modeling~\cite{cosinebg2}, are used to set Gaussian priors for the background. The systematic uncertainties are included in the fit as nuisance parameters with Gaussian priors. 

Prior to the data fit, the fitter was tested with simulated event samples. Each simulated dataset is prepared by Poisson random extraction of the modeled background spectrum, assuming a background-only hypothesis. Marginalization to obtain the posterior probability density function (PDF) for each simulated sample is performed to set the 90\,\% confidence level upper limits. The 1,000 simulated experiments result in $\pm$1$\sigma$ and $\pm$2$\sigma$ bands of the expected median sensitivity.

\begin{figure*}[!htb]
\begin{center}
\begin{tabular}{cc}
\includegraphics[width=0.49\textwidth]{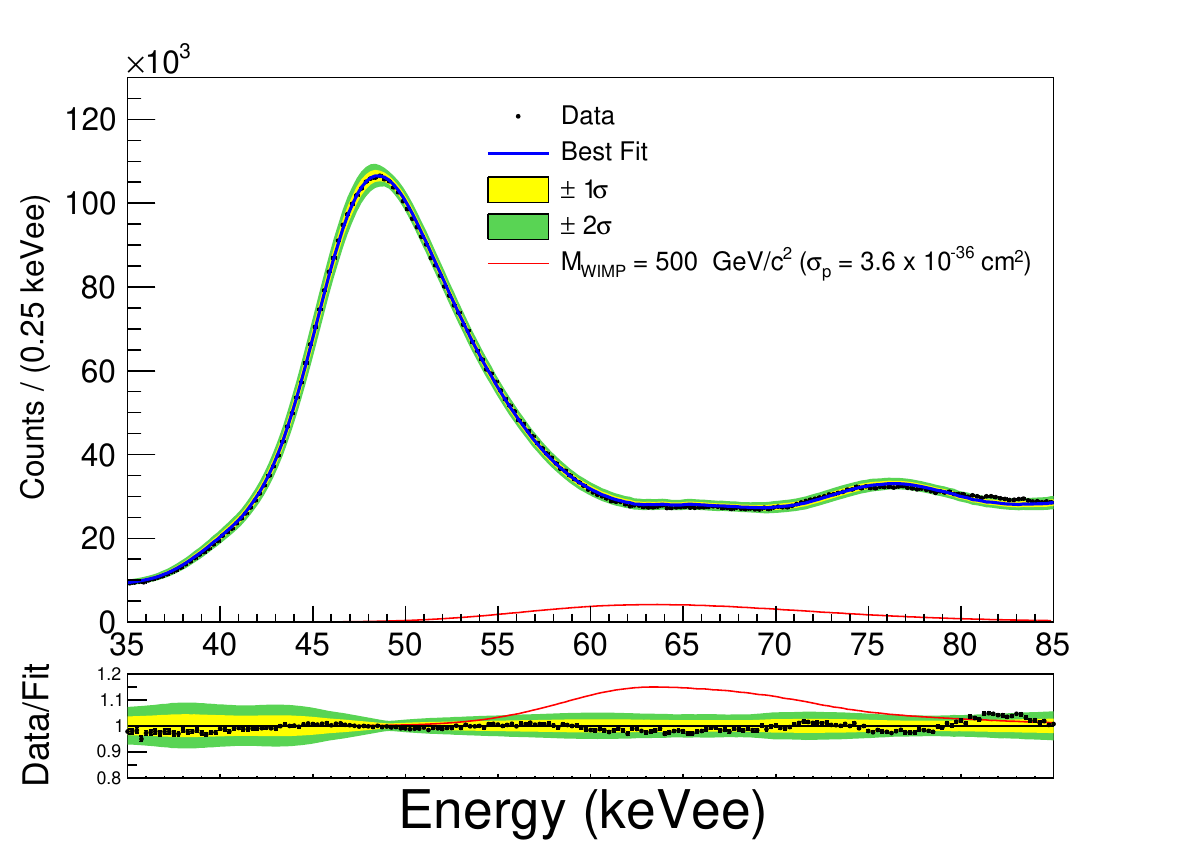} &
\includegraphics[width=0.49\textwidth]{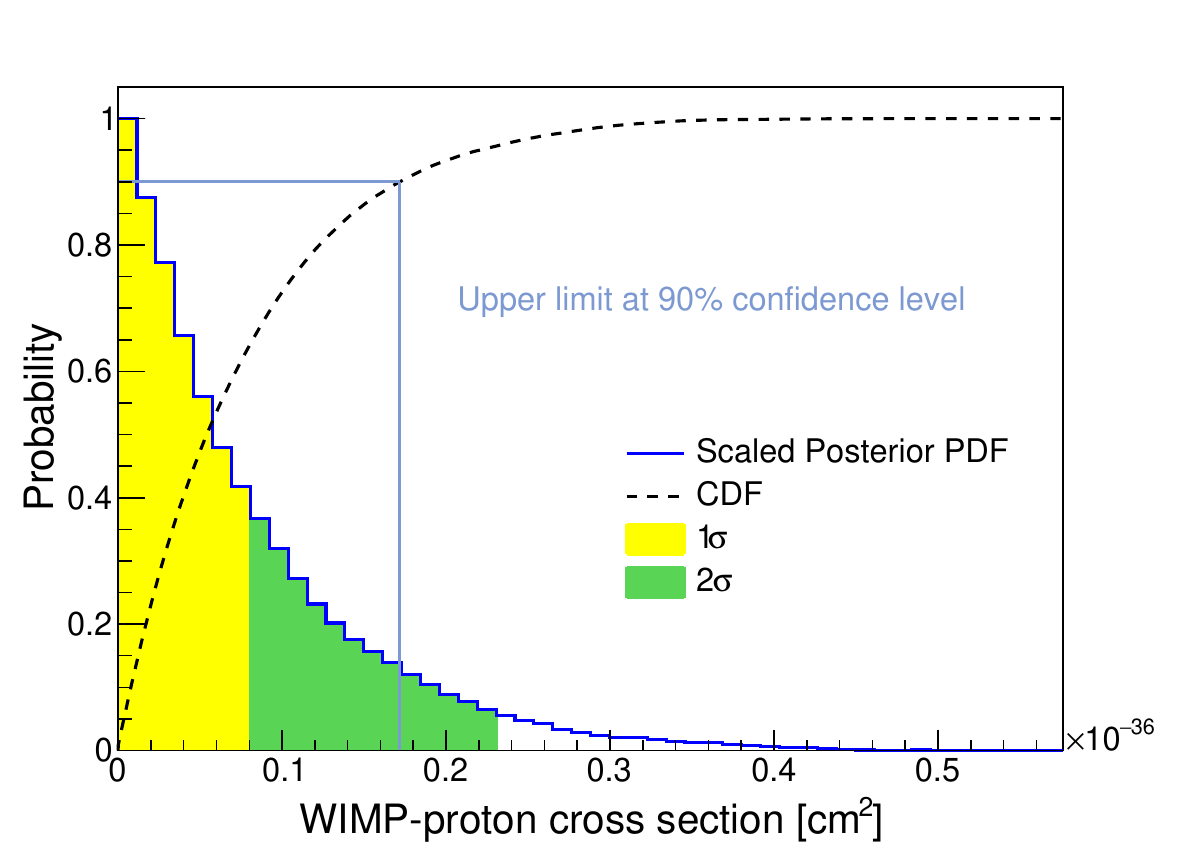}\\
(a) & (b) \\
\end{tabular}
\caption{(a) The black-filled circles represent the data points showing the summed energy spectra from the five crystals.  The solid blue line indicates the result of the fit with a 500\,GeV/c$^2$ WIMP mass signal. 
  		The expected signal shape of the 500\,GeV/c$^{2}$ WIMP mass is a red solid line, assuming WIMP-proton spin-dependent cross sections of $3.6\times10^{-36}$\,cm$^{2}$, which is 30 times higher than the 90\,\% confidence level upper limit. 
			(b) It is an example of the posterior probability density function (PDF) and cumulated density function (CDF) for the 1.7-year of COSINE-100 data and a WIMP mass of 500\,GeV/c$^{2}$. The posterior PDF is scaled so that  the maximum value is one. In this PDF, the best fit ({\it i.e.} the most probable cross section) points to a null signal.  Therefore, we set the 90\,\% confidence level upper limit. 
 The exclusion limit at a 90\,\% confidence level is obtained from the CDF matched with 0.9. The yellow and green areas represent the 1\,$\sigma$ and 2\,$\sigma$ confidence intervals, respectively.  }
\label{fig:FitResult}
\end{center}
\end{figure*}

Fits to the data are performed for each of the 12 WIMP masses considered in the same way as the simulated data. As an example, the data fit with a WIMP mass of 500~${\rm GeV/c^{2}}$ is presented in Fig.~\ref{fig:FitResult}. 
The summed event spectrum for the five crystals is shown in Fig.~\ref{fig:FitResult} (a), which corresponds to the null observation.
For comparison, the expected signal for the WIMP mass 500\,GeV/c$^2$ with spin-dependent cross section of $3.6\times10^{-36}$cm$^2$ is overlaid. 
This cross section is 30 times higher than the measured 90\,\% confidence level upper limit. 
Figure~\ref{fig:FitResult} (b) shows the posterior PDF and its cumulative distribution function  of data for this example.

\begin{figure}[!htb]
\begin{center}
\includegraphics[width=0.49\textwidth]{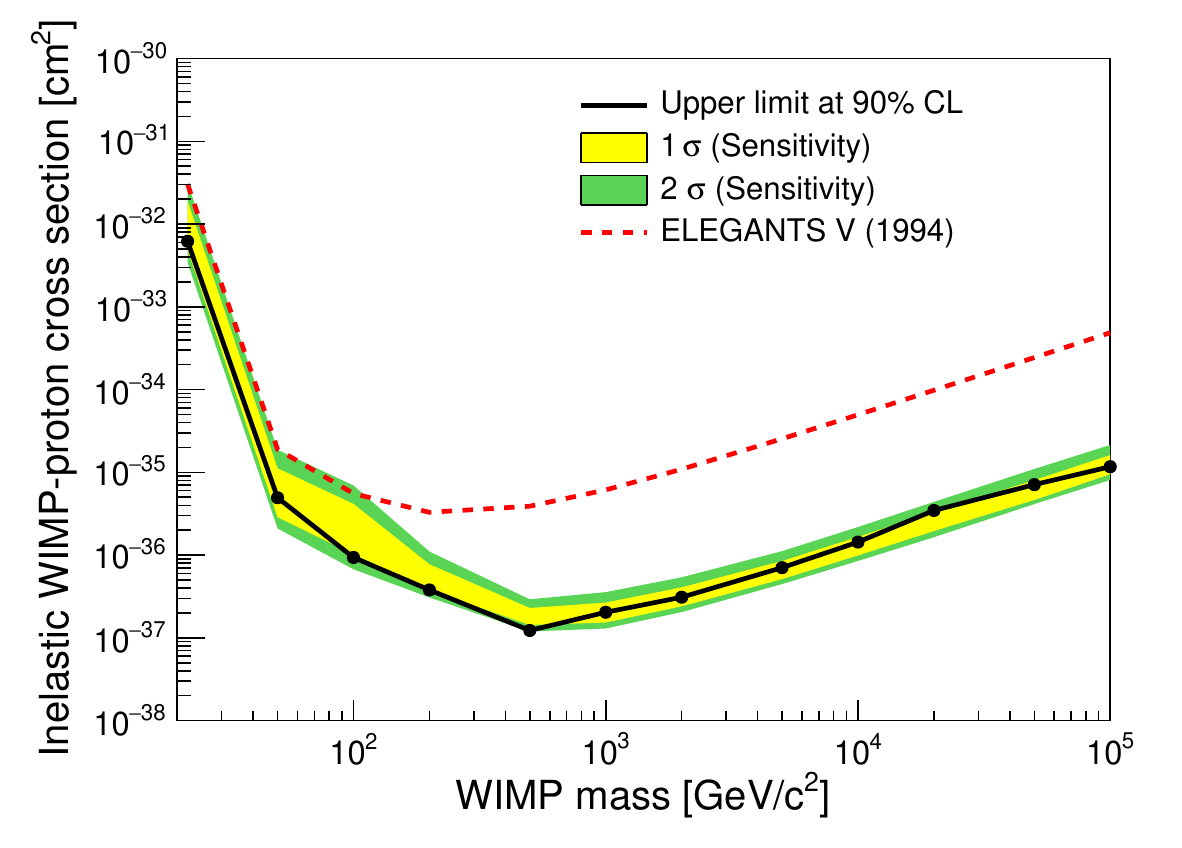} 
\caption{The observed 90\,\% confidence level exclusion limits on the WIMP-proton spin-dependent cross section of WIMP-\I127 inelastic interactions from the 1.7-year of COSINE-100 data are shown, along  with the $\pm$1\,$\sigma$ and $\pm$2\,$\sigma$ bands for the expected sensitivity under the background-only hypothesis. The limits are compared with a WIMP interpretation of the ELEGANTS-V upper limit of 57.6\,keVee event rate for the WIMP-\I127 inelastic scattering hypothesis. }
\label{fig:sensitivity}
\end{center}
\end{figure}

No excess of events that could be attributed to WIMP-\I127 inelastic interactions is found in the 12 different WIMP masses considered. 
The posterior probabilities of the signal were consistent with zero in all cases, and 90\% confidence level limits are determined, as an example shown in Fig.~\ref{fig:FitResult}~(b). 
Figure~\ref{fig:sensitivity} shows the 90\% confidence level exclusion limits from the COSINE-100 data with $\pm$1$\sigma$ and $\pm$2$\sigma$ expected bands of exclusion limits from the simulated experiments.

For comparison, we interpret a 90\,\% confidence level upper limit of 
ELEGANTS-V from the inelastic scattering event rate of 9.8$\times$10$^{-2}$\,counts/kg/day~\cite{FUSHIMI1994400} as the WIMP-\I127 inelastic scattering cross section, as shown in Fig.~\ref{fig:sensitivity}. 
Although extracted event rates depend on the WIMP mass, we assume the same event rates of the ELEGANTS-V interpretations considering similar shapes of signal spectra due to the dominant deexcitation energy of 57.6\,keV.
Our results improve the exclusion limit by order of magnitude from the previous search for the same channel and are the most stringent result to date  in the spin-dependent WIMP-proton cross section via the inelastic scattering channel.

Because of the dominant background from $^{210}$Pb of 46.5\,keV $\gamma$ with Auger electrons to ROI, the inelastic  WIMP-\I127 scattering signal for the low-mass WIMP is encompassed with the $^{210}$Pb background as one can see in Fig.~\ref{fig:sigMC}. 
It is also affected by the systematic uncertainty of the $^{210}$Pb modeling, especially from the energy scale that increases uncertainties of the event rates near the signal region and increases the fluctuation of the expected limit bands, as one can see in Fig.~\ref{fig:sensitivity}. Compared with the ELEGANTS-V interpretation, which assumes flat background, the low-mass WIMP limit  from this work reflects the influence of the the $^{210}$Pb background.  

Our R\&D program for the development of low-background NaI(Tl) crystal detectors has resulted in NaI(Tl) detectors with significantly reduced $^{210}$Pb background~\cite{COSINE:2020egt, 10.3389/fphy.2023.1142765}, which will be applied to the COSINE-200 experiment~\cite{Ko:2022pmu}. With the realization of the COSINE-200 experiment, the sensitivities to search for the WIMP-\I127 inelastic interaction will be enhanced. 

\section{Conclusion}
We performed a search for WIMP-\I127 inelastic scattering events from 57.6\,keV deexcitation $\gamma$ with the nuclear recoil in the 1.7 years COSINE-100 data. 
The single-hit energy spectrum was fitted with signal and background models in the energy range of 35--85\,keVee. 
We found no evidence of the WIMP-\I127 inelastic interaction signals, allowing us to set 90\,\% confidence level exclusion limits on the WIMP-proton spin-dependent interaction cross section.  The best limit is  1.2$\times$10$^{-37}$cm$^2$ of a WIMP mass of 500\,GeV/c$^2$. It is the most stringent limit for the spin-dependent WIMP-proton interaction using WIMP-nucleus inelastic scattering process. 

\acknowledgments
We thank the Korea Hydro and Nuclear Power (KHNP) Company for providing underground laboratory space at Yangyang and the IBS Research Solution Center (RSC) for providing high performance computing resources. 
This work is supported by:  the Institute for Basic Science (IBS) under project code IBS-R016-A1, NRF-2019R1C1C1005073, NRF-2021R1A2C3010989 and NRF-2021R1A2C1013761, Republic of Korea;
NSF Grants No. PHY-1913742, DGE-1122492, WIPAC, the Wisconsin Alumni Research Foundation, United States; 
STFC Grant ST/N000277/1 and ST/K001337/1, United Kingdom;
Grant No. 2021/06743-1, 2022/12002-7 and 2022/13293-5 FAPESP, CAPES Finance Code 001, CNPq 303122/2020-0, Brazil.
\bibliographystyle{PRTitle}
\providecommand{\href}[2]{#2}\begingroup\raggedright\endgroup
\end{document}